\documentclass[%
 reprint,
 amsmath,amssymb,
 aps,
]{revtex4-2}

\usepackage{graphicx}
\usepackage{dcolumn}
\usepackage{bm}

\begin{document}

\preprint{APS/123-QED}

\title{Adaptation to extreme stress under \\ the growth-survival fitness trade-off}

\author{Nandita Chaturvedi}
 \email{cnandita@ncbs.res.in}

\author{Charuhansini Tvishamayi}%
\author{Shashi Thutupalli}
\affiliation{%
Simons Center for the Study of Living Machines, National Center for Biological Sciences (TIFR), Bangalore, India
}%


\begin{abstract}
Microbial adaptation to extreme stress, such as starvation, antimicrobial exposure, or freezing often reveals fundamental trade-offs between survival and proliferation. Understanding how populations navigate these trade-offs in fluctuating environments remains a central challenge. We develop a quantitative model to investigate the adaptation of populations of yeast (\textit{Saccharomyces cerevisiae}) subjected to cycles of growth and extreme freeze-thaw stress, focusing on the role of quiescence as a mediator of survival. Our model links key life-history traits—growth rate, lag time, quiescence probability, and stress survival—to a single underlying phenotype, motivated by the role of intracellular trehalose in the adaptation of yeast to freeze-thaw stress. Through stochastic population simulations and analytical calculation of the long-term growth rate, we identify the evolutionary attractors of the system. We find that the strength of the growth-survival trade-off depends critically on environmental parameters, such as the duration of the growth phase. Crucially, our analysis reveals that populations optimized for growth-stress cycles can maintain viability alongside growth-optimized populations even in the absence of stress. This demonstrates that underlying physiological trade-offs do not necessarily translate into fitness trade-offs at the population level, providing general insights into the complex interplay between environmental fluctuations, physiological constraints, and evolutionary dynamics.
\end{abstract}

\maketitle
\section{Introduction}

Adaptation to extreme stress in microbial systems often occurs at a cost to their ability to grow quickly under ambient conditions \cite{wenger2011hunger, zakrzewska2011genome}. Specifically, an increase in survival under oxidative stress, glucose limitation, freeze-thaw, and heat exposure redirects resources from growth pathways to those dedicated to damage reduction and repair \cite{shabestary2024phenotypic}. While trade-offs are not universal across all environmental niches \cite{bennett2007experimental, buckling2007experimental}, they are a common outcome of adaptation to extreme stressors. These evolutionary trade-offs directly influence the emergence of generalists or specialists depending upon the extremity and frequency of environmental change \cite{lambros2021emerging, lopez2008tuning, tikhonov2020model}. Fitness trade-offs can manifest in distinct life history traits by affecting growth rates, cellular response to starvation, survival and characteristics of the lag phase in distinct ways. This leads to a complex interplay between environmental variation, population dynamics and underlying physiological constraints during the process of evolution. The switch to quiescence is a crucial part of the population response to starvation, and its role in evolutionary adaptation remains relatively unexplored, perhaps with an exception in the case of bacterial persistor cells \cite{sagot2019quiescence}.

Here we present a model of adaptation of \textit{Saccharomyces cerevisiae} to repeated cycles of growth and nutrient exhaustion followed by freeze-thaw stress, based upon experiments covered in previous work \cite{tvishamayi2024convergent}. Evolution under these conditions produces a clear fitness trade-off: freeze-thaw survival increases while the growth rate in ambient conditions decreases, and additionally the duration of lag time decreases. Our model simulates the evolution of a yeast population through these repeated cycles and explicitly addresses the roles of quiescence and lag time in the adaptation process.

The evolved cells in our experimental system are smaller, denser and lighter than the wild type cells and show higher intracellular trehalose concentration in stationary phase \cite{tvishamayi2024convergent}. Trehalose, a dimer of glucose plays a  crucial role both in entry to and exit from quiescence, as well as cellular response to extreme stress. Trehalose has been observed to bind to the cell wall and protect protein molecules which prevents damage from freezing or desiccation \cite{olsson2020structural, aranda2004trehalose, tapia2015increasing, arguelles2000physiological, panek1963function, diniz1999preservation}. Total intracellular trehalose shows correlation with the density of individual cells both in stationary phase cultures and during continuous growth. Immediately upon exit from quiescence, cells metabolize trehalose to re-enter the cell cycle. Cells that lack trehalose initiate growth more slowly and frequently exhibit poor survivability \cite{shi2010trehalose}. Thus, cells with more trehalose exhibit a shorter lag phase when re-introduced to rich media \cite{opalek2022ecological}. A shorter lag phase has also been linked to a higher adaptability to changing environments \cite{chu2016lag}. This establishes a dual trade-off: not only between growth rate and stress survival, but also between growth rate and lag time, a phenomenon previously observed in \textit{E. coli} \cite{basan2020universal}.

Further, when yeast is subject to cycles of growth and freeze-thaw stress without allowing the population to reach stationary phase, the population crashes, suggesting that quiescence acts as an important step to the evolution of freeze-thaw tolerance. Recent work in yeast has shown that stress response often follows a general strategy rather than evolving to specific stress conditions \cite{wenger2011hunger, zakrzewska2011genome, stern2007genome}. For instance, yeast evolved under aerobic glucose limitation shows few trade-offs with growth in other carbon-limited environments \cite{zakrzewska2011genome}. While the fitness of the evolved strain increases in comparison to yeast growing in rich media, these fitness improvements can arise from multiple mutational paths. This general stress response is most evident at the level of the phenotype and includes producing higher levels of stationary phase trehalose, thickening of the cell wall, a reduction in size \cite{tvishamayi2024convergent}, and is underpinned by large-scale, genome-wide transcriptional reprogramming \cite{stern2007genome, soontorngun2017reprogramming}. Therefore, while our model is grounded in a specific experimental evolution system, its results can be generalized to different microbial populations evolving under cycles of growth and stress exposure. 

We highlight the role of quiescence in adapting to extreme stress. Yeast populations undergo a diauxic shift following glucose depletion, where cells reduce their growth rate to readjust their metabolism for the subsequent phase of slow, respiratory growth. After nonfermentable carbon sources have been consumed, the population enters stationary phase. Stationary phase cultures are complex and heterogeneous, composed of a large fraction of quiescent, long-lived, largely daughter and young mother cells, and a nonquiescent subpopulation of cells \cite{de2012essence, sagot2019cell, breeden2022quiescence, aragon2008characterization, opalek2022ecological}. The non-quiescent subpopulation eventually perishes under conditions of long term starvation. Quiescence is a reversible exit from the cell cycle in the stationary phase, distinct from growth arrest due to nutrient starvation  \cite{de2012essence, daignan2011proliferation, valcourt2012staying}.

Thus, quiescence of yeast cells in conditions of starvation is a milder stress response which could prime them for harsher conditions, enabling survival during long periods of starvation \cite{herman2002stationary}. Quiescent cells share several key characteristics with response to extreme stress, including smaller, denser cells with thicker cell walls \cite{allen2006isolation, aragon2008characterization}, and also show increased resistance to diverse environmental stresses \cite{herman2002stationary, werner1993stationary, alarcon2011quiescence}. However, the exact relationship between quiescence and canonical cellular stress response remains unclear \cite{werner1993stationary, sagot2019cell}. Distinct from stress-resistant cells, quiescent cells show cell division arrest and stochastic switching between the cell cycle and quiescence. The dynamics of entry into and exit from quiescence has been shown to be a noisy process that couples deterministic memory effects of preceding cell cycle with stochastic switching \cite{wang2017exit}. Yet, the large scale rearrangement of gene expression, metabolic state and cellular processes that takes place in quiescence may show similarity to stress response in yeast \cite{sun2021cellular, shabestary2024phenotypic, kaplan2021observation, woronoff2020metabolic}.

We therefore identify intracellular trehalose produced in stationary phase as a primary phenotypic marker. We posit that this intracellular trehalose concentration determines the population's growth rate, survival probability,  lag time, and each cell's likelihood of going into quiescence. Our model integrates growth, lag and survival-mediated selection with the stochastic switch to quiescence.  

In Section \ref{sec_trehalose_model} we introduce our model of microbial evolution under growth-starvation-stress cycles that treats the switch to quiescence as a stepping stone to stress-resistance. The model qualitatively captures the evolution of yeast populations under cycles of growth, starvation, and freeze-thaw stress. We specifically confirm its predictions for the evolutionary trajectory of the central phenotype, the amount of intracellular trehalose in stationary phase, with experimental results. Using analytic calculations, we estimate the long term growth rate of a population under growth-stress cycles in Section \ref{sec_lgr} and use it to calculate the optimum phenotype for different evolution programs. We find that despite the growth-survival trade-off, populations evolved under growth-stress cycles can have comparable fitness to those those evolved in the absence of stress. This suggests the possibility of coexistence. We explore this suggestion using simulations of finite resource competitive growth of the two optimal strains.

\section{Trehalose Mediated Quiescence, Lag Time, Growth and Survival \label{sec_trehalose_model}}
\begin{figure*}%
    \centering    
  \includegraphics[width=0.8\linewidth]{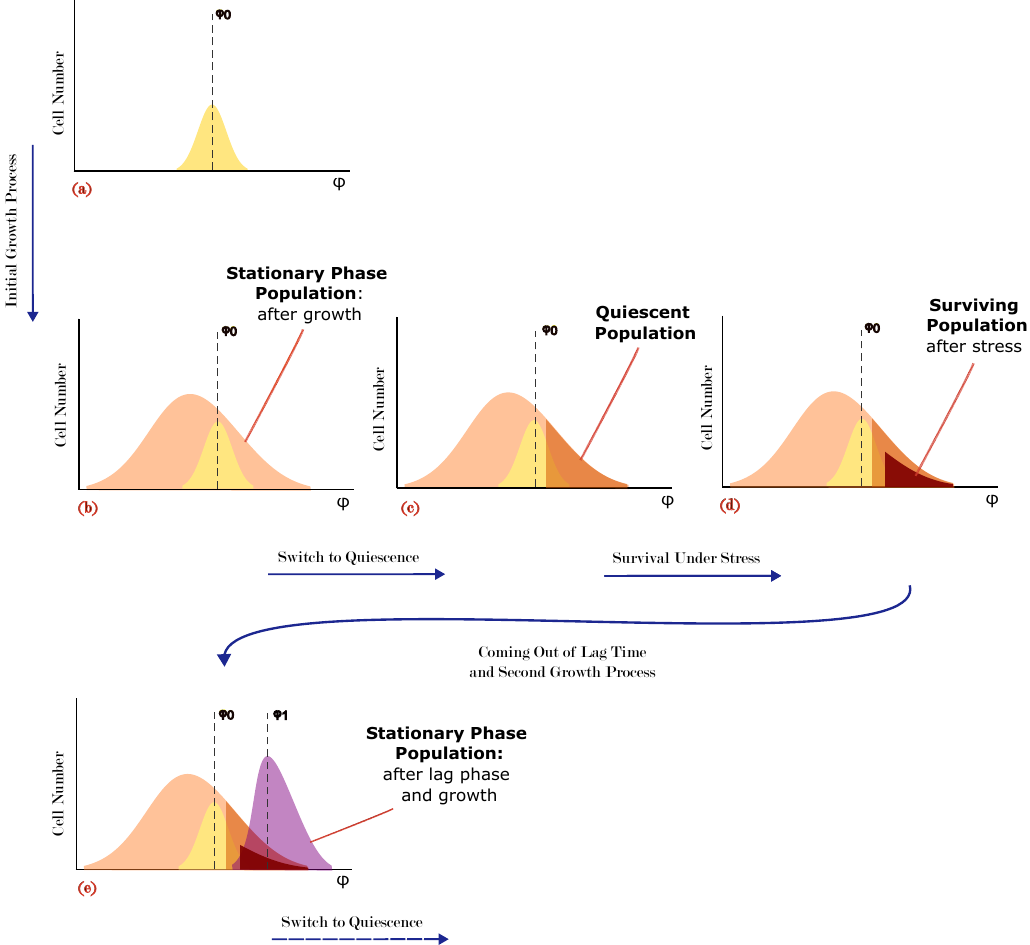}
   \caption{ Schematic showing the first round of the evolution simulation. (a) An initial population drawn from a Gaussian distribution in $\phi$ is selected and allowed to undergo resource limited growth till it reaches a stationary phase at resource exhaustion. (b) The stationary phase population distribution is broader than the initial distribution due to mutations, and its mean is shifted towards the left due to higher growth rates at lower $\phi$ values. (c) This population then differentiates into quiescent and non-quiescent cells, with the probability of quiescence imposing a lower threshold in $\phi$ on the population distribution. (d) The population is subsequently exposed to extreme stress and a smaller fraction of the quiescent population survives above a second threshold value of $\phi$ imposed by the survival probability. (e) As the population is given a resource source again, cells in the distribution exit lag phase and begin growth. Higher $\phi$ values mean lower lag times, but also lower growth rates. The next stationary phase population is a result of these two competing effects.  The simulation then continues with the population again differentiating into quiescent and non-quiescent cells.  \label{fig_schematic}}
\end{figure*}
We highlight a few features of the experiments in \cite{tvishamayi2024convergent} to motivate our model: (1) The survival fraction for yeast undergoing several rounds of growth followed by freeze-thaw stress shows a gradual increase from about $2\%$ to $70\%$ over $25$ cycles. Since the increase is gradual, this indicates that the adaptation to freeze-thaw is not through a single sweeping mutation or rearrangement of biological pathways, but rather through a series of changes, either genetic or otherwise, that allow for incremental fitness increases. (2) The population shows survival and subsequent adaptation when freeze-thaw is applied after it reaches stationary phase. If the population is frozen in its exponential phase, it collapses within a few cycles and no survival is seen. (3) The evolved strain of yeast after about $25$ cycles is characterized by a shortened lag phase after stress, a slightly lower growth rate and higher levels of trehalose production in stationary phase. (4) The cells of the evolved strain are smaller, lighter but denser, and show less budding, similar to key characteristics of quiescent cells.

Entry into quiescence is a stochastic process, while the efficiency of a cell's exit from quiescence is expressed in the duration of the lag phase upon re-entry into rich media. The duration of the lag phase is also determined by the amount of intracellular trehalose produced in stationary phase, as trehalose is preferentially metabolized to re-enter the cell cycle. Further, higher trehalose levels correspond to lower cellular damage under physical stress, as well as higher survival fractions under freezing and desiccation. Lastly, the cellular mechanisms that produce trehalose as part of a broader and general stress response show a trade off with the pathways that promote cell division. Thus, as trehalose production under stressful conditions increases, growth rate of the cell upon reentry into the growth phase decreases. All of this is observed in our experimental results, and we postulate that the stationary phase trehalose levels in a cell can determine the cell's growth rate in exponential phase, quiescence probability, survival probability, and lag time upon reentry into the cell cycle. 

For building our model and simulations, we assume that a single phenotype, $\phi$ is able to capture different aspects of a cell's fitness, including growth rate, probability of making a switch into quiescence in stationary phase, survival probability in extreme stress and the lag time of the cell once reintroduced into fresh media. This primary phenotype corresponds to the amount of intracellular trehalose in the cell when the population enters stationary phase.

We start with a discrete population of $N$ individual cells whose phenotype, given by $\phi$, is drawn initially from a continuous Gaussian distribution with mean $\phi_0$ and a narrow peak with standard deviation $\sigma_\phi^0$. Note that each cell has a phenotype given by $\phi$, and the population is characterized by a distribution in $\phi$. This is shown in Figure \ref{fig_schematic}(a).

Our simulation is divided into four phases; (1) growth phase, (2) quiescence phase, (3) survival phase, and (4) lag phase. To mimic the experimental setup, the first phase in our simulation is a growth phase where the population is introduced into fixed resource, $r_0$. Each time step in this phase of the simulation is a new generation of the population, and we assume non-overlapping generations. Each cell in the population has a different growth rate determined by its phenotype. We assume the dependence of growth rate (or fecundity per unit time in the simulation) on $\phi$ to be quadratic,
\begin{equation}
    g(\phi)=g_0(1-\phi/\tau_g)^2 \label{eq_gr}
\end{equation}

where $g_0$ is the highest achievable growth rate and $\tau_g$ the amount of trehalose that inhibits growth to effectively zero. Note that although our population does not typically reach $\phi$ values close to $\tau_g$ during the course of the simulation, we set $g(\phi)=0$ for $\phi>\tau_g$. 

At each generation time step in the growth process, we calculate the number of offspring for each cell using $g(\phi)$. We use a quadratic function for $g(\phi)$ to allow the growth rate to fall faster with increasing $\phi$ at lower values and slower as $\phi$ reaches higher values. Since the number of cells is discrete while the growth function and phenotype are continuous functions of $\phi$, we apply a ceiling function to $g(\phi)$ to calculate the number of offspring for each cell add them to the next population generation. 

Each offspring is subject to a small mutation, which acts to change the phenotype, $\phi$ of the offspring with respect to its parent. This mutational effect on offspring phenotype is drawn from a Pearson Distribution at a mutation rate $m$ for each offspring. The Pearson Distribution is a continuous probability distribution with a long tail defined by its kurtosis. Our distribution has a mean at zero, mutational standard deviation $\sigma_m$, a kurtosis $k_m$ and no skew. A kurtosis $k=3$ is a Normal distribution, with outliers becoming more probable with increasing kurtosis. For our simulations we use $m=0.8$ and $k_m=5$. Since the distribution of mutational effects has no skew, the median of the distribution is at $0$. This means that most mutations have no effect upon the offspring's phenotype, preserving the value inherited from the parent cell. Simultaneously, because of our choice of a high kurtosis, some large effect mutations, both positive and negative are possible. Although the mutations almost never drive $\phi$ to be a negative value, we explicitly reject mutations that cause $\phi<0$ in our simulations.

For each cell division event, we assume a constant amount of the finite resource, $r_d$, is consumed. The growth process continues till the resource runs out and the population faces resource exhaustion. At this stage, as shown in Figure \ref{fig_schematic}(b) the population distribution is broader because of the effect of mutations, and shifted towards lower $\phi$ because of the growth rate function when compared to the initial population. For our simulations we use $r_d=0.5$ and $r_0=10^5$. 

It should be noted that although we consider $\phi$ for each cell corresponds to trehalose concentrations produced by the cell eventually in the stationary phase, it also determines its quiescence probability, probability of survival and lag time. Thus, our model implies that mutations accrued in growth process can bias the cell's performance in other life history stages, although no trehalose is actually produced in the growth phase.

Following experimental observations, upon reaching starvation conditions we assume the stationary phase population differentiates into a heterogeneous population of quiescent and non-quiescent cells in the next simulation phase. This occurs in a single simulation time step. Our model considers the switch to quiescence as a stochastic process, where the probability of each cell in stationary phase going into quiescence is given by
\begin{equation}
    p_q(\phi)=\frac{1}{(1+\exp^{-(\phi-\tau_q)/\gamma_q})},
    \label{eq:pq}
\end{equation}

where $\tau_q$ is the value of $\phi$ at which probability of quiescence is $50\%$ and $\gamma_q$ is a parameter quantifying how quickly the probability rises to $1$ as $\phi$ increases. The functional form, as well as the $\tau_q$ values we chose allows for probability of quiescence to gradually increase with $\phi$ and saturate at a high value. For each cell we decide the fate of the cell by comparing this probability with a random number drawn from a uniform probability distribution. If the probability given by Equation \ref{eq:pq} is higher than the random number the cell goes into a quiescent state. The effect of this probability of quiescence is to impose a lower threshold in $\phi$ on the population, as shown in Figure \ref{fig_schematic}(c).

\begin{figure}%
    \centering    
  {\includegraphics[width=\linewidth]{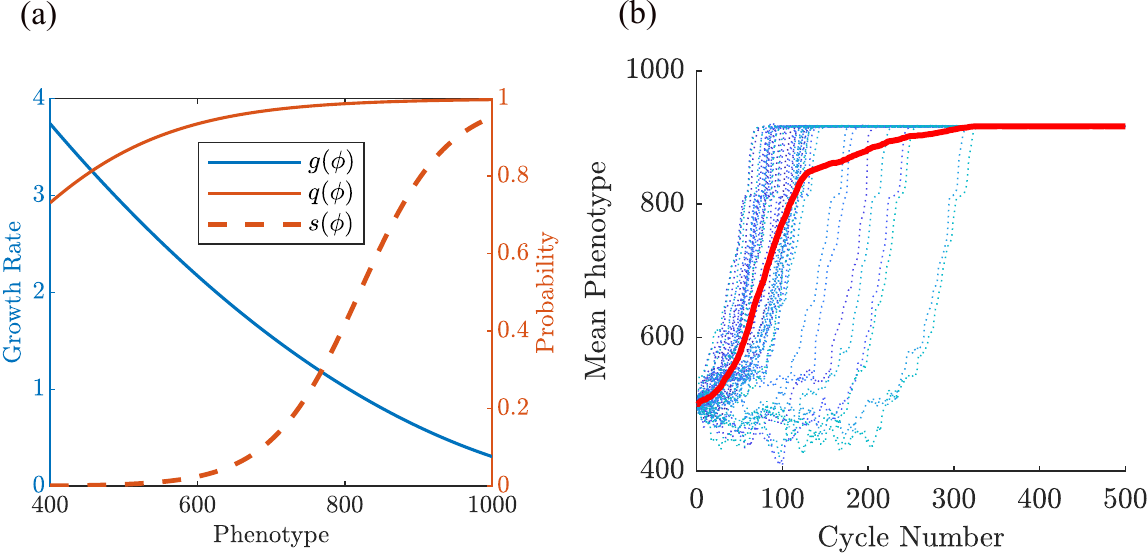}}%
    \caption{ (a) Growth rate, quiescence probability and survival probability as functions of the phenotype. (b) The mean phenotype of the population first increases stochastically before settling to an equilibrium value. The red curve is the average over $100$ replicate populations, and the blue trajectories represent individual replicates. As can be seen, while the trajectory is stochastic, the final equilibrium phenotype remains the same across replicates. Here $m=0.8$, $\sigma_m=3$, $k_m=5$, $l_{0}=30$, $\tau_l=1000$, $g_0=40$, $\tau_g=1200$, $r_0=10^5$, $\tau_q=400$, $\gamma_q=100$, $\tau_s=850$, $\gamma_q=50$.\label{fig_trehalosemodel}}
\end{figure}

Entry into stationary phase is followed by exposure to extreme stress, which in the experimental system corresponds to freeze-thaw. This is the third phase in our simulation, consisting of a single simulation time step update. For the purpose of our model we assume that the non-quiescent population entirely perishes upon exposure to extreme stress. Among the quiescent population, cells with higher $\phi$ have a higher probability of survival. We model the probability of survival,

\begin{equation}
    p_s(\phi)=\frac{1}{(1+\exp^{-(\phi-\tau_s)/\gamma_s})},
    \label{eq_sp}
\end{equation}

where $\tau_s$ is the $\phi$ value at which probability of survival under freeze thaw is $50\%$ and $\gamma_s$ is a parameter quantifying how quickly the probability rises to $1$ as $\phi$ increases. Again, we choose $\gamma_s$ and $\tau_s$ such that the probability of survival is close to zero for low values of $\phi$ and then increases relatively quickly in an interval given by $\gamma_s$, mimicking a switch-like transition to having enough trehalose production in each cell to lend the cell resistance to freeze-thaw. In general we expect that $\gamma_s<\gamma_q$ and $\tau_s>\tau_q$ since the stress is harsher to cells than nutrient scarcity in stationary phase. Thus, both quiescence and the freeze thaw process shift the mean phenotype of the population to higher values since cells with lower phenotype amounts perish. This is shown in Figure \ref{fig_schematic}(d). The functions for growth rate, quiescence probability and survival probability are shown in Figure \ref{fig_trehalosemodel}(a).

We model the dilution process before re-introduction into fresh resource of the experimental setup by randomly selecting a fraction $d$ of the population. This can introduce additional stochasticity into the evolution process. 

For the final phase of the simulation, when the surviving population is re-introduced into a fresh growth medium, we assume the population remains initially in lag phase, with the lag time of each cell determined by its phenotype. We model cells with higher trehalose concentrations to have a shorter lag time, and resume their growth earlier. We assume the simplest possible functional form, a linear relationship between the lag time and $\phi$,

\begin{equation}
    l(\phi)=l_0(1-\phi/\tau_l),
    \label{eq_lt}
\end{equation}

where $l_0$ represents the maximum lag time possible, and $\tau_l$ the $\phi$ value at which the lag time reduces to zero. We set $l(\phi)=0$ for all $\phi>\tau_l$ for the sake of our simulation, although $\tau_l$ is set large enough that $\phi$ does not exceed it in the simulations. Cells also consume a smaller amount of the resource, $r_l=0.1$ while they are non-growing in lag phase. Longer lag times thus mean less resource available for growth and the total amount consumed before growth depends on the length of the lag phase. A larger phenotype value thus means a shorter lag phase, and more time available for growth, while simultaneously causing a lower growth rate. The next generation to reach stationary phase, shown in Figure \ref{fig_schematic}(e), is the result of competing effects caused by a phenotype dependent lag time.

Figure \ref{fig_trehalosemodel}(b) shows the gradual increase in the mean phenotype of the population as it evolves to stress exposure. When the population starts growth after exit from lag phase with lower values of mean $\phi$, the effect of a decreasing growth rate is not sufficient to stop the increase in mean $\phi$ of the population caused by a shorter lag phase. In the growth phase, cells with higher $\phi$ are able to produce a larger number of offspring and take over the population. This explains the initial increase in the mean phenotype.

Each time the population undergoes a complete round of growth and nutrient exhaustion, we have a resulting heterogeneous population, with cells with higher $\phi$ having a higher probability of entering quiescence. Thus, the distribution of quiescent cells now has a higher mean $\phi$ than the entire stationary population taken together. The cells that are able to survive are a larger proportion of the entire population because of this upward shift in $\phi$. Hence, our simulation includes both a component of phenotype-dependent stochastic switching to quiescence, as well as growth, lag and survival mediated selection for phenotype values. The trade-off between the decrease in growth rate and the positive effect on lag time and survival from increasing trehalose sets the final equilibrium value of the phenotype in the simulations. The effect of a lower growth rate begins to dominate at larger phenotype values, causing the mean phenotype of the population to stagnate.

\begin{figure}%
    \centering    
  \includegraphics[width=\linewidth]{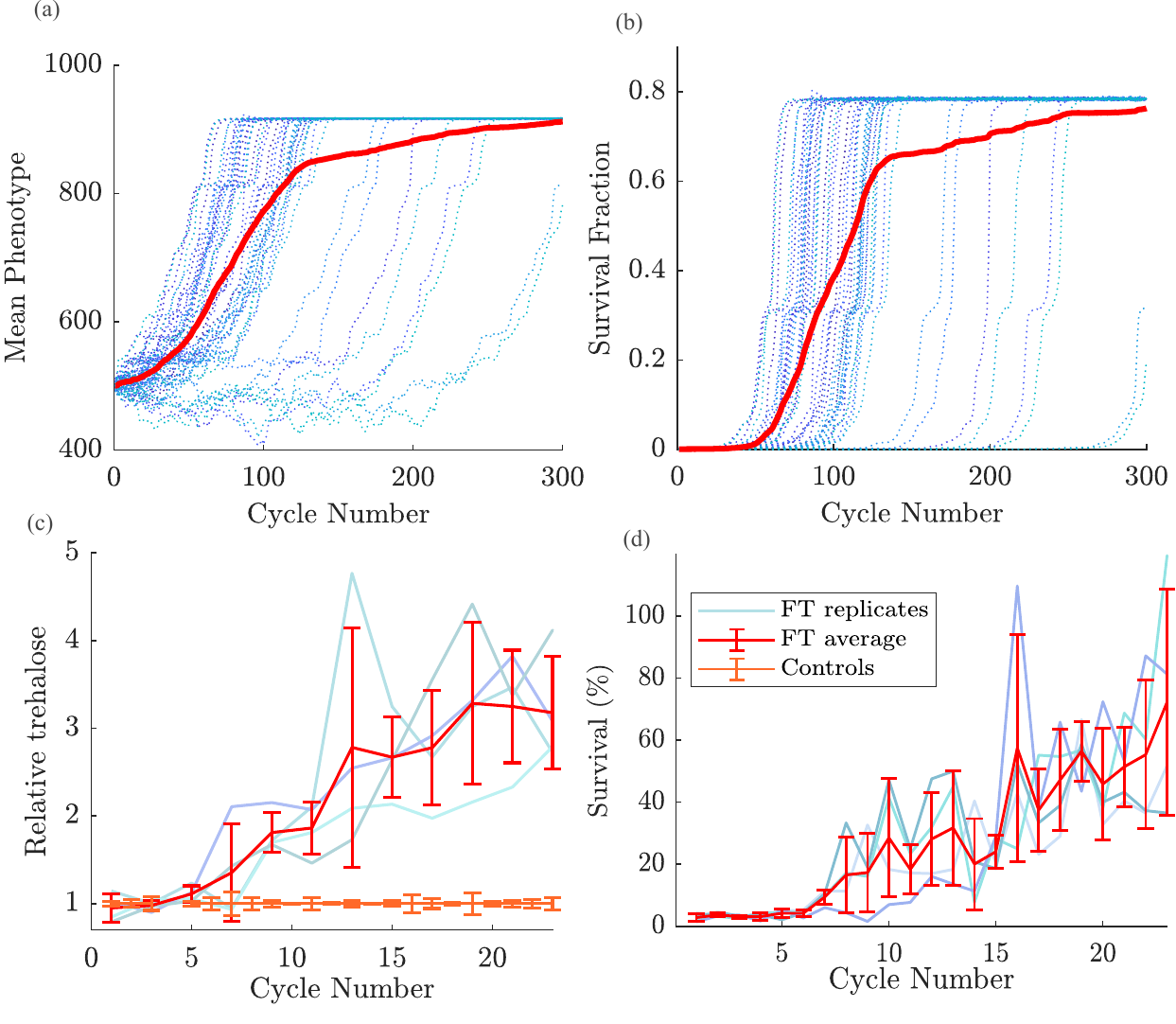}
    \caption{(a) Mean phenotype of the population and (b) survival fraction with cycle number for $100$ replicate populations. For comparison, (c) is the mean trehalose amount at stationary phase in the yeast populations as a function of cycle number. Four replicates and their average is shown. (d) Survival fraction of yeast populations under freeze-thaw stress increases from about $2\%$ to $70\%$ in as little as 25 cycles of freeze-thaw. Figures (c) and (d) are reproduced from \cite{tvishamayi2024convergent}. \label{fig_experiment}}
\end{figure}

We see an increase in the phenotype and survival as the evolution process progresses, before settling at an equilibrium value. This is shown in Figure \ref{fig_experiment}. The blue curves represent individual replicate populations, while the red curve represents the average over $100$ replicates. The behavior of populations is stochastic in the early stages of the evolutionary process, due to small numbers of cells that survive freeze-thaw stress, but the equilibrium value they settle to remains conserved across all replicates. 

We compare Figure \ref{fig_experiment} (a) and (b) to the trehalose evolution curve (c) and the survival fraction curve (d) obtained from the experimental evolution of yeast under freeze thaw. Both sets of curves show qualitative agreement. The rapid increase in the survival rate is caused by the large tailed mutational distribution we use which biases evolution towards large jumps. We note also the delay in increase of both the survival curves, which arises, in the case of our simulation, from the sigmoidal dependence of the survival probability upon trehalose concentration.

\begin{figure}%
    \centering    
  {\includegraphics[width=\linewidth]{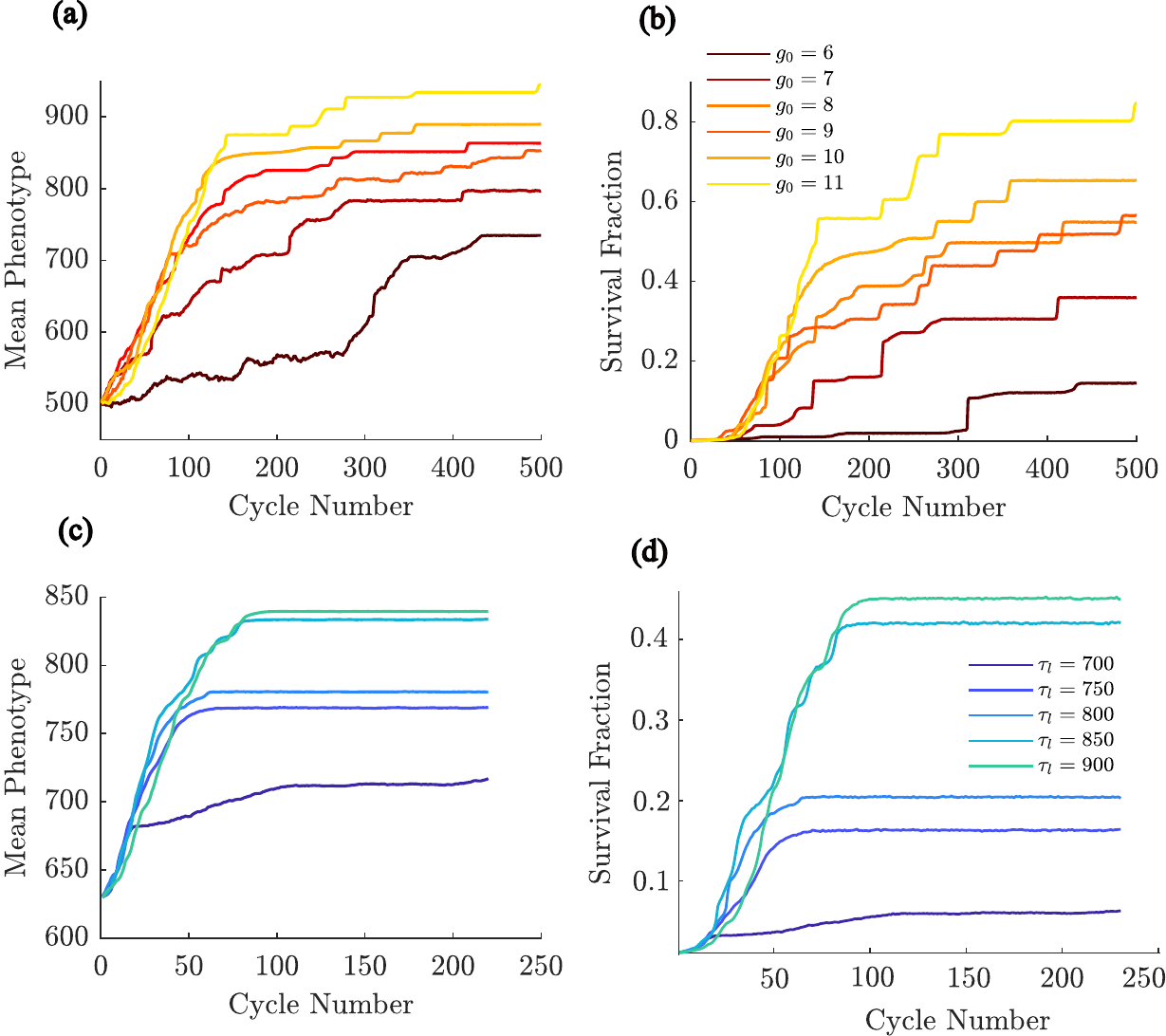}}%
    \caption{Mean phenotype and survival fraction for changing $g_0$ (figure (a) and (b)) and $\tau_l$ (figure (c) and (d)). Each line represents the mean value taken over $40$ replicates. As can be seen in (a) and (b), increasing $g_0$ increases final equilibrium phenotype as well as survival. Two competing effects come into play, a higher $g_0$ means a larger population level variation at each time point for selection to occur. It also means that the total time for growth reduces if the total resource is kept constant. However, increasing $g_0$ implies also a higher selection from growth. The effect of decreasing growth time dominates. For (a) and (b), $m=0.8$, $\sigma_m=3$, $k_m=5$, $l_{0}=30$, $\tau_l=1000$, $r_0=10^5$, $\tau_g=1200$, $\tau_q=400$, $\gamma_q=100$, $\tau_s=850$, $\gamma_q=50$. In (c) and (d) we can see increasing the phenotype value at which lag effectively drops to zero also increases the final phenotype and survival. This is because increasing $\tau_l$ means that the lag time is higher at the same phenotype. Since this leaves less time for growth, the contribution of selection from growth decreases, allowing the equilibrium value of the phenotype as well as survival to climb higher. For (c) and (d), $m=0.8$, $\sigma_m=3$, $k_m=5$, $l_{0}=30$, $r_0=10^5$, $g_0=20$, $\tau_g=1200$, $\tau_g=1200$, $\tau_q=400$, $\gamma_q=100$, $\tau_s=850$, $\gamma_q=50$. \label{fig_param}}
\end{figure}

To determine the effect of model parameters on equilibrium values of the phenotype and survival fraction, we first vary $g_0$.  This is shown in Figure \ref{fig_param} (a) and (b). Increasing $g_0$ increases the number of offspring at each time point, leading to the resource being exhausted faster, thus less time is spent by the population in its growth phase. This causes the effect of selection from growth to decrease, allowing the phenotype to increase to higher values and biasing the population towards maximizing survival. Thus, we see an increase in both the equilibrium value of the phenotype as well as survival fraction as $g_0$ increases.

Similarly, Figure \ref{fig_param} (c) and (d) show that decreasing the rate at which the lag time decreases as a function of the phenotype, controlled by increasing $\tau_l$, also leads to an increase in the equilibrium values of the phenotype and survival fraction. We find that increasing the value of $\tau_l$ leads to the lag time increasing at each value of the phenotype. This means that a larger proportion of the resource is consumed by non-growing cells, leading to the population having less time in the exponential growth phase before the resource is exhausted. 

Changing the mutation rate and variance of mutation effects only affect the dynamics of the climb to equilibrium, leaving the equilibrium value of the phenotype unaffected. 

Thus, we find that the amount of time spent in the growth phase, the maximum growth rate, among other parameters can affect the equilibrium value of the phenotype and survival by changing the strength of the trade-off. Note that this hints towards the sensitivity of the final mean phenotype of the population to the experimental setup. We write down the exact form of this dependence in Section \ref{sec_lgr} by estimating the long term growth rate of the population.

\section{Long Term Growth Rate and Comparing Fitness \label{sec_lgr}}

We calculate the long term growth rate of the evolving population to help us to analytically recover some of the results of our model. Since the life cycle of microbes such as yeast consists of several life history stages, a single trait such as growth rate or survival cannot be taken as a complete indicator of strain fitness. Fitness is a complex combination of these traits, and the long term growth rate can effectively capture this. 

Consider a population of cells evolving under growth-stress cycles given by the distribution $n(\phi, t)$. As in the last section, the growth rate, probability of survival under stress and lag time all depend on the phenotype of each cell. Through the growth phase, the number of cells with phenotype $\phi$ at time $(t+1)$ is given by $g(\phi) n(\phi,t)$, where $g(\phi)$ is the growth rate at phenotype $\phi$.  If the population is allowed to grow for $T_g$ time till the resource runs out, at the end of the growth phase the number of cells at each phenotype is 

\begin{equation}
n(\phi,T_g)=g(\phi)^{T_g-t_l(\phi)} n(\phi,0),
\end{equation}

where $t_l(\phi)$ is the lag time of a cell with phenotype $\phi$. We measure time in units of generation time, so $g(\phi)$ is only a numerical factor.

The population is then exposed to extreme stress, such as freeze-thaw, and the fraction of cells at each phenotype that survive the stress is $s(\phi)$. Thus after one cycle of lag-growth-starvation-stress and dilution, the population has a distribution

\begin{equation}
n(\phi,T_g+T_s)=d s(\phi)g(\phi)^{T_g-t_l(\phi)} n(\phi,0).
\end{equation}

where $T_s$ is the duration of stress exposure and $d$ is the dilution rate. Note that we do not explicitly account for transition to quiescence here for simplicity of calculation, which means that $s(\phi)$ does not act upon a smaller quiescent population, but rather on the entire population. The survival process in this calculation can be thought of as an effective process that combines quiescence and survival from our simulations. After $c$ cycles we have a population distribution
\begin{equation}
n(\phi,c(T_g+T_s))=[d s(\phi)g(\phi)^{T_g-t_l(\phi)}]^c n(\phi,0).
\end{equation}

Thus, the subpopulation with phenotype $\phi$ grows exponentially as the number of cycles tends to infinity if $d s(\phi)g(\phi)^{T-t_l(\phi)}>1$ and dies out exponentially if $d s(\phi)g(\phi)^{T-t_l(\phi)}<1$. At large cycle numbers, the growth rate of the population is given by $ \Lambda=\frac{1}{C}\log{\frac{n(\phi,C(T_g+T_s))}{n(\phi,0)}}$. Thus, the effective long term growth rate of the subpopulation with phenotype $\phi$ under growth-stress cycles can then be written as

\begin{equation}
    \Lambda_s=\log{ [d s(\phi)g(\phi)^{T_g-t_l(\phi)}]} \label{eq_lgr}
\end{equation}

At long times, the phenotype that maximizes the long term growth rate dominates the population. Thus, finding the $\phi$ values at which $\Lambda_S$ is maximum will give us the optimum phenotype for a population evolved under growth-stress cycles. Further, the rate at which the $\Lambda_s$ falls off from this maxima can give us an estimate of the variation in the population. 

\begin{figure}%
    \centering    
  {\includegraphics[width=\linewidth]{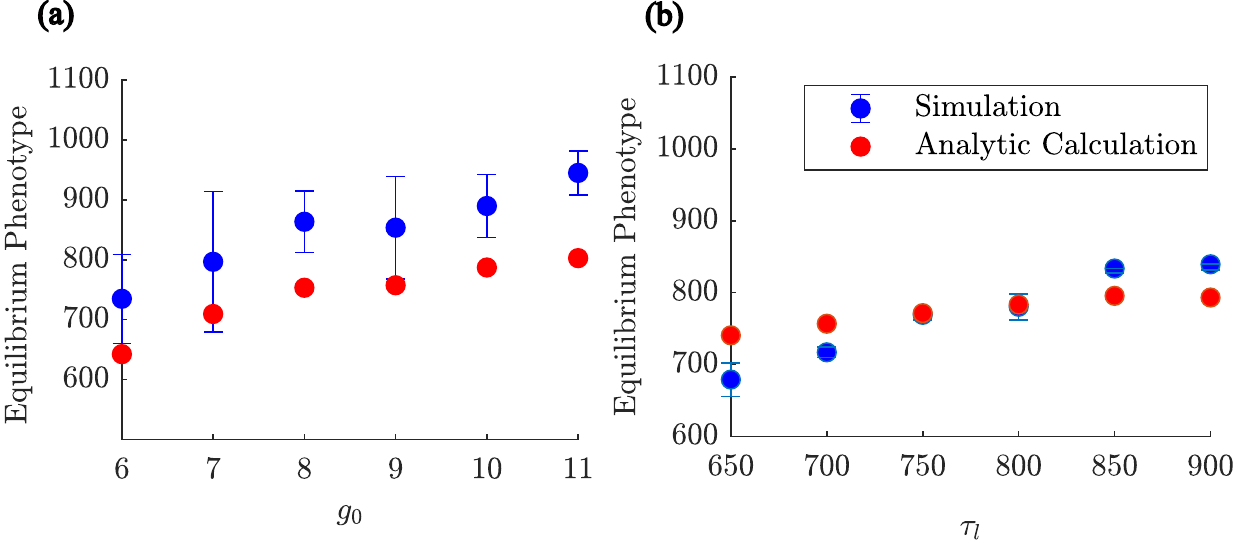}}%
    \caption{We compare the equilibrium values of the population's mean phenotype (in blue) with the optimum phenotype $\phi$ that maximizes equation \ref{eq_lgr} (red) for different values of $g_0$ and $\tau_l$. Each point from the simulations represents the mean and standard deviation from $20$ different replicates. The difference between the two curves can be explained by the assumption that the growth time remains constant during the duration of evolution. We use the mean value of $T$ from simulations to determine the optimum phenotype analytically. This leads to an difference of the equilibrium phenotype value from the analytically predicted value. To see how the growth time varies with cycle number, see Supplementary Information. Unless otherwise specified, here we use $m=0.8$, $\sigma_m=3$, $k_m=5$, $l_{0}=30$, $\tau_l=1000$, $r_0=10^5$, $g_0=20$, $\tau_g=1200$, $\tau_g=1200$, $\tau_q=400$, $\gamma_q=100$, $\tau_s=850$, $\gamma_q=50$. \label{fig_compare}}
\end{figure}

Figure \ref{fig_compare} shows a comparison of the equilibrium phenotype values from the simulation with the optimal phenotype that maximized the long term growth rate in equation \ref{eq_lgr}. As was seen in the last section, the equilibrium phenotype increases with both an increase in $g_0$ and $\tau_l$. Note that we made the assumption above that the total time of growth, $T$, remains constant across cycles. This assumption does not hold in the simulations as the time of growth decreases with increasing population size as the evolution progresses forward. Growth time in our simulations drops from a relatively high value in the beginning to a lower equilibrium value. For the growth time dynamics, see Supplementary Information. We use the mean of the growth time from simulations as $T$ in \ref{eq_lgr} to calculate the optimum phenotype. Since we are comparing a finite resource growth process (in the simulations) to a finite time growth process (in the analytic calculation), some difference in the two predictions is seen. However, the analytical prediction agrees with the values from the simulation to a large degree.

\begin{figure*}%
    \centering
    {{\includegraphics[width=0.8\linewidth]{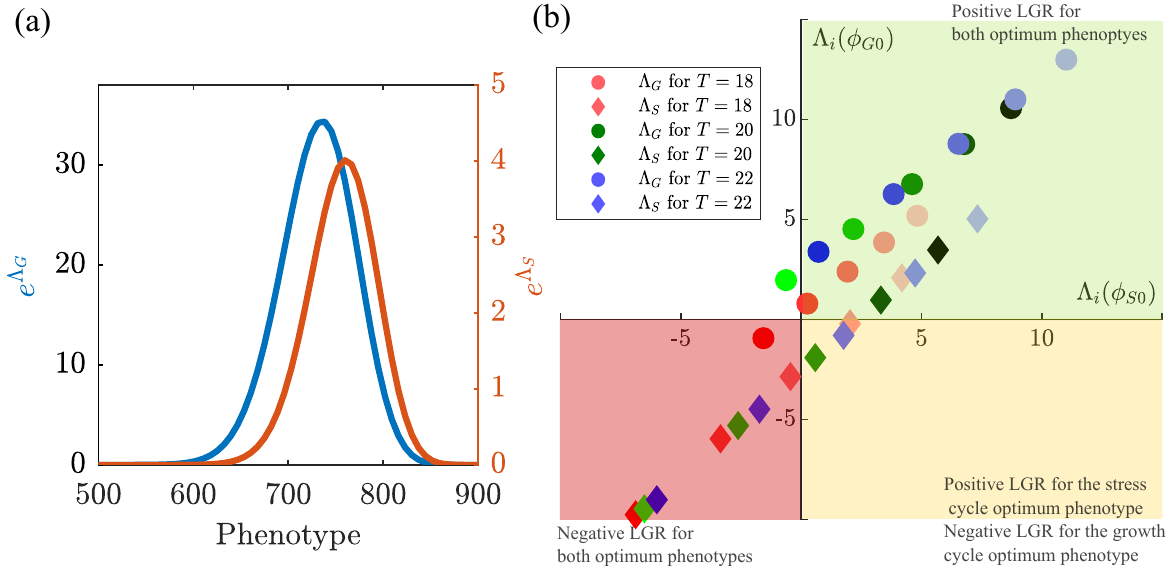} }}%
    \qquad
    \caption{(a) A plot of $e^{\Lambda_S}$ and  $e^{\Lambda_G}$ shows that the value of $\phi$ at which they are maximized, as well as their values at the optimum are different. (b) The optimal phenotype in each experimental setup (with and without stress) was calculated for different values of $T$ and $g_0$. Red, blue and green series show the three values of $T$ chosen while $g_0$ was chosen to be $6, 8, 10, 12$ and $14$ for each $T$ value (increasing as you go from lower left to upper right in each series). The long term growth rate of these 'growth type' and 'survival type' optimal strains with phenotypes $\phi_{G0}$ and $\phi_{S0}$, respectively, was then calculated with and without stress growth cycles. The same color circle and diamond show the performance of $\phi_{G0}$ and $\phi_{S0}$ evolved with the same $g_0$ and $T$ with and without stress, so can be compared to each other. Note that when the long term growth rate is negative, the strain is expected to go extinct. As can be seen, for most parameter pairs, the performance of the population remains in the upper right quadrant for under growth cycles (circular points), \textit{i.e.} populations of both $\phi_{G0}$ and $\phi_{S0}$ grow well under only growth cycles. When each point is compared with its counterpart under stress cycles (diamond points), you can see a shift towards lower $\Lambda$ for both phenotypes.  Some points remain in the upper right quadrant, implying both $\phi_{G0}$ or $\phi_{S0}$ are able to grow under stress, only with reduced rates (with $\Lambda_i(\phi_{S0})$ typically decreasing less than $\Lambda_i(\phi_{G0})$. Some points are shifted to the lower left quadrant, implying neither $\phi_{G0}$ or $\phi_{S0}$ are able to grow under stress. A few points are in the lower right quadrant, showing survival of $\phi_{S0}$ under stress conditions and extinction of $\phi_{G0}$. In other words, a strain evolved under stressful conditions survive under stress cycles, while that under growth cycles do not. \label{fig_lgr}}
\end{figure*}

If the population undergoes cycles of growth, starvation and dilution without exposure to stress, we can write the long term growth rate as $ \Lambda_g=\log{[d \space g(\phi)^{T_g-t_l(\phi)}]}$. Figure \ref{fig_lgr} (a) shows $e^{\Lambda_g}=d g(\phi)^{T-t_l(\phi)}$ and $e^{\Lambda_S}$ as a function of $\phi$. Since $s(\phi)$ increases with $\phi$, while $t_l(\phi)$ and $g(\phi)$ decrease with $\phi$, both functions have a unique maximum. The position of this maximum is different for the case with and without stress, representing the varying degrees of the growth-survival trade-off.

The long term growth rate is an effective tool to compare two populations evolved under different conditions; under solely growth cycles and under growth-stress cycles. We calculate the value of the phenotype which maximizes the long term growth rate for each evolution program, $\phi_{G0}$ and $\phi_{S0}$, respectively. We then compare $\Lambda_G$ and $\Lambda_S$ of both optimal phenotypes in the two cases. This is shown in Figure \ref{fig_lgr}. The three different point color schemes, blue, green and red, represent three different values of $T$, while the different shades in each color represent optimum phenotypes for different $g_0$ values, with $g_0$ increasing from lower left to upper right quadrant. The same color circle and diamond show the performance of $\phi_{G0}$ and $\phi_{S0}$ evolved with the same $g_0$ and $T$ with and without stress respectively. 

When the long term growth rate is negative, the strain is expected to go extinct quickly. For most parameter pairs, the performance of the population remains in the upper right quadrant under growth cycles (circular points), \textit{i.e.} populations of both $\phi_{G0}$ and $\phi_{S0}$ grow well under only growth cycles. This is striking since it shows that although $\phi_{S0}$ is evolved to be optimum under growth-stress cycles, it can perform well under solely growth cycles. Further,  $\phi_{G0}$ and $\phi_{S0}$ show comparable long term growth rates, which points to the possibility of coexistence. Thus, despite the strong growth-survival trade-off, the overall fitness does not show a strictly inhibiting trade-off for $\phi_{S0}$ under growth conditions. When each circular point is compared to its counterpart in growth-stress cycles (diamond points), a shift is observed towards lower $\Lambda$ for both phenotypes. 

Some populations, with larger $g_0$ values, remain in the upper right quadrant, implying both $\phi_{G0}$ or $\phi_{S0}$ are also able to grow with positive long term growth rates under under growth-stress cycles, only with reduced rates (with $\Lambda_i(\phi_{S0})$ typically decreasing less than $\Lambda_i(\phi_{G0})$. Thus, both evolved populations of $\phi_{G0}$ and $\phi_{S0}$ are generalists with different emphasis on growth and survival. 

A few populations under growth-stress fall in the second quadrant, showing survival of $\phi_{S0}$ under stress conditions and extinction of $\phi_{G0}$. In other words, a strain evolved under stressful conditions survives under growth-stress cycles, while one evolved under growth cycles does not and is a specialist. However, as mentioned,  the counterparts of these populations under solely growth cycles exist in the upper right quadrant. This shows that even though a population of $\phi_{S0}$ evolves to exist under stress conditions when $\phi_{G0}$ cannot, under solely growth cycles the two populations have comparable long term growth rates and may coexist. This further suggests that despite the growth-survival trade-off, populations evolved under growth-stress may not show significant trade-offs with their fitness under only growth cycles.

Some populations shift to the lower left quadrant in Figure \ref{fig_lgr} under growth-stress cycles, implying neither $\phi_{G0}$ or $\phi_{S0}$ are able to grow under stress. This occurs at lower values of $g_0$, where the low fecundity does not allow the population to recover its numbers after the significant decrease in its population size upon exposure to stress. Increasing both $g_0$ and $T$ can help the population recover and reduce this effect. 

We note that while here we are considering a single optimum phenotype, in real world situations we expect to see a population with a distribution in phenotypes. The width of this population distribution is set by the balance between the variance in fitness mutation effects and the effective selection strength. Thus, if the population width is large enough to accommodate different phenotypes and hence different long term growth rates, coexistence of two strains with similar positive long term growth rates may be possible. This could happen in two ways. 1) Under conditions where one of the strains is optimal, but the population distribution is broad enough to accommodate the second. This is made possible by an increase in the variance of mutational effects, $\sigma_m$.  2) Under conditions when both strains are suboptimal and hence have comparable fitness. Since both strains can be generalists as the populations in the first quadrant of Figure \ref{eq_lgr}, this means that in cases where the emphasis on lag time and growth rate is different than the conditions of evolution, coexistence may be possible. The $x$ and $y$ axis of Figure \ref{eq_lgr} represent the fitness of each strain for an environment with the same growth time, $T$, at which the strains are evolved to maximize fitness. 

We test the second possibility by simulating the competitive growth of the two optimal strains with a common finite resource. We calculate the optimal phenotypes $\phi_{G0}$ and $\phi_{S0}$ using the parameter values used to generate Figure \ref{fig_lgr}, with $g_0=40$ and $T=22$. We further calculate their respective growth rates, survival probabilities and lag times using equations \ref{eq_gr}. \ref{eq_sp} and \ref{eq_lt}. Let $N_i$ (where $i=(1,2)$) be the biomass of strain \textit{i} at each simulation time step. Our simulations start with equal and small biomass values of two strains (taken to be $N_i=0.1$) in a finite resource, $r_0$. At the beginning of the simulation, both strains are assumed to be in lag phase. At each time step we test whether both populations continue to be in their lag phase. While a population is in lag phase, its biomass does not increase but the resource is depleted by a small amount proportional to the biomass. Once either strain crosses its calculated lag time we assume it follows the Monod equation for effective growth rate in a finite resource,
\begin{equation}
    g_{i}=g_i^0 \frac{r}{K_i+r},
\end{equation}
where $i=(1,2)$, $g_i^0$ is the maximum growth rate of strain \textit{i} calculated using Equation \ref{eq_gr} and the constant $K_i$ is the value of resource amount $r$ at which the growth rate of strain \textit{i} falls to half its maximum value. At each time step of the simulation we update the biomass of both strains by adding $g_i N_i$, as well as the resource which reduces by an amount $\delta r_i$ (where $i=(1,2)$) per unit biomass of each strain produced. We allow each strain to grow till the amount of resource per unit biomass of the strain falls below a predefined threshold ($r/N_i<10$), at which point the strain enters stationary phase. In stationary phase, growth is halted while the resource continues to be consumed at a smaller rate $0.1\delta r_i$, like in the lag phase. Note that the two strains enter both growth and stationary phases independently. Once the resource value drops below a second threshold ($r/N_i<5$), we put both strains either through stress by multiplying each strain's biomass by its survival probability, or through dilution, by multiplying each strain's biomass by $d=0.3$. This process is then repeated by replenishing the resource amount once again to $r_0$. We consider a strain extinct once its biomass falls to one-thousandth its initial value ($N_i<0.0001$).

The value of growth time $T$ in the evolution simulations in the previous section is set by the amount of initial resource added, $r_0$, although the exact relation between them is difficult to determine analytically. Hence it is not directly possible to know the value of $r_0$ for which our two strains are optimal. However, we do know that changing $r_0$ leads to a change upon the emphasis on growth and stress in the cycles. 

\begin{figure}%
    \centering    
  {\includegraphics[width=\linewidth]{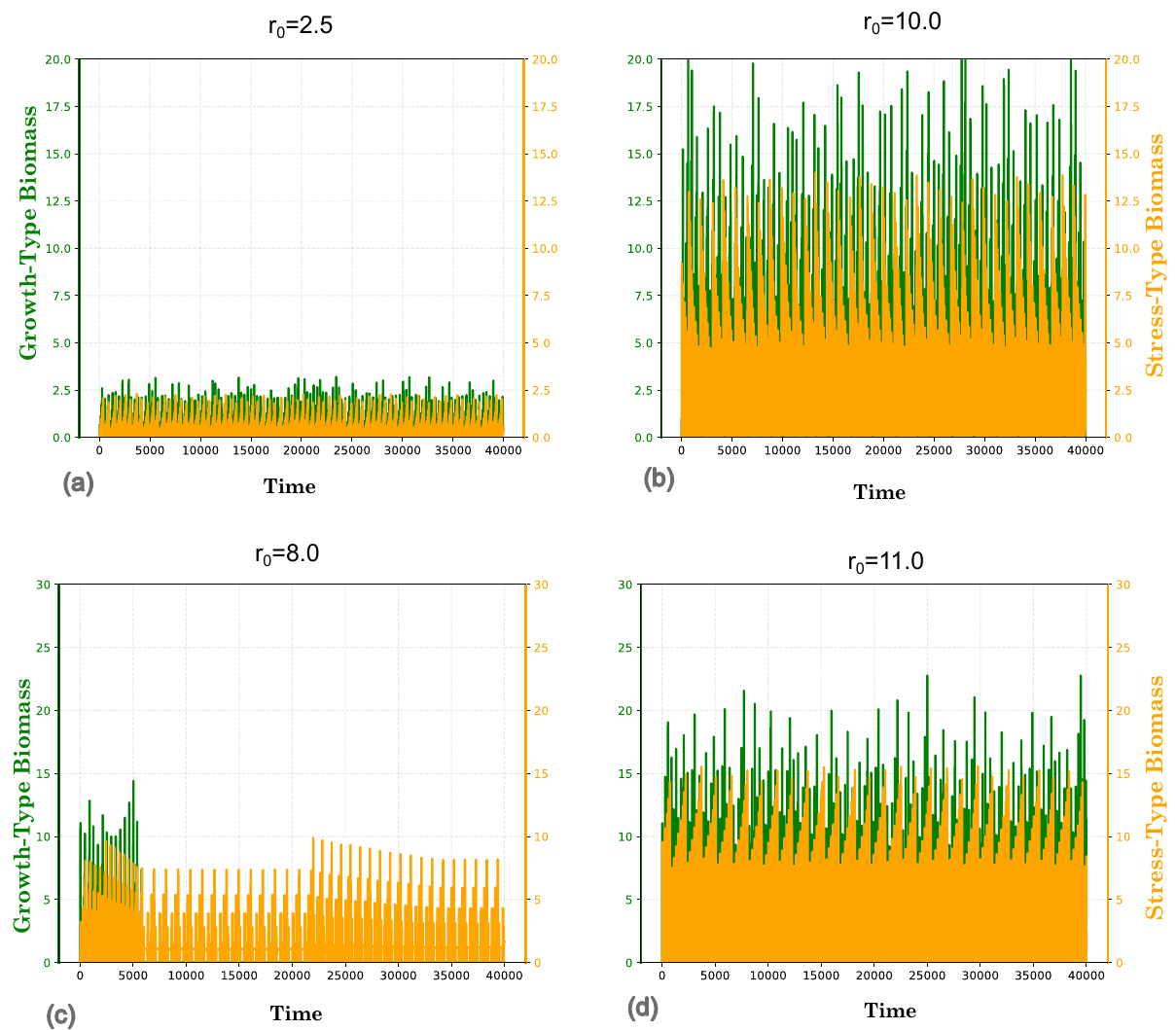}}%
    \caption{We test the possibility of coexistence of growth-type and stress-type strains in growth cycles (a-b) and growth-stress cycles (c-d).  We first determine the optimal phenotype for both conditions using Equation \ref{eq_lgr} and its corresponding growth rate, lag time and survival probability. We then run resource limited growth competition simulations with two strains of the derived characteristics. (a) When the simulation has solely growth cycles, we see both strains survive for most resource values, $r_0=2.5$ is shown here. The growth-type strain has a higher biomass. (b) A higher value of $r_0=10.0$ with growth only cycles is shown here. (c) Starting with $r_0=8.0$ in growth-stress cycles, we find that the growth type strain goes extinct, while the stress type strain survives. (d)  Starting with $r_0=11.0$  and growth-stress cycles, we find that that the two strains are able to coexist. For these results we use $l_{0}=30$, $\tau_l=1000$, $r_0=10^5$, $g_0=30$, $\tau_g=1200$, $\tau_g=1200$, $\gamma_q=100$, $\tau_s=850$ which gives us $\phi_{S0}=465.81$, $\phi_{G0}=390.74$, $g_{G0} =13.6437$, $
g_{S0}=11.2300$, 
$t_{l,G0}=18.28$,
$t_{l,S0}=16.03$, $s_{G0}= 0.0001$ and $s_{S0}= 0.0005$. The parameters for the coexistence simulation are $K=3$, $\delta r_{G0}=0.29$, $\delta r_{S0}=0.4$, the threshold for $r/N_i$ for transition into stationary phase is $10$.  Both strains are started from $N_i=0.1$ and the extinction threshold is $N_i=0.0001$. \label{fig_coexist}}
\end{figure}

Figures \ref{fig_coexist} (a) and (b)  shows two biomass time series for two different values of $r_0$ for growth only cycles. We find that at resource values across a range, including $r_0=2.5$ and $10.0$, we find that coexistence between the two occurs, with the maximum biomass of the stress-type strain being lower than that of the growth-type strain. Here the stress-type strain is at a disadvantage because of a lower growth rate and larger lag time. Yet because of comparable positive long term growth rates (the upper right quadrant of Figure \ref{fig_lgr}) the stress-type strain does not go extinct up to $40,000$ simulation time steps.

Figures \ref{fig_coexist} (c) and (d) shows similar time series plots for growth-stress cycles. At different resource values we are able to see (c) a case where only the stress-type strain survives, and (d) a case where both are able to coexist. Out-competition in (c) occurs because the long term growth rate of the growth-type strain is driven small or negative due to the inclusion of survival probability (the lower right quadrant of Figure \ref{fig_lgr}). Upon increasing $r_0$ to $11$, we effectively increase the long term growth rate of both strainFigure \ref{sec_lgr}. It is interesting to note that the maximum biomass of the growth-type strain is still higher than that of the stress-type even though its long term growth rate is similar. This is because the growth rate of the growth-type strain is higher than that of the stress-type strain, while their long term growth rate remains similar. Thus, the amount of biomass both are able to produce in the first exponential phase, which is larger for the growth-type strain, remains maintained through subsequent cycles. 

\section{Discussion}

Growth-survival trade-offs, observed across microbial species, help explain key ecological phenomena like co-existence, population heterogeneity, and the preference for resource-scarce environments \cite{zhu2024shaping, bruggeman2023trade}. We present a model of adaptation to extreme stress under such a trade-off. To our knowledge, this is a novel effort to model the role of quiescence in evolutionary adaptation to extreme stress. This approach is supported by evidence that stress response in yeast may follow a general strategy rather than evolving to specific stress conditions \cite{wenger2011hunger, stern2007genome, jahn2017adaptive}. Studies in bacteria suggest that exposure to lower concentrations of antibiotics may prepare cells to take on the challenge of lethal concentrations \cite{hughes2012selection,sandegren2014selection}. Further, the existence of the growth-lag trade-off suggests that starvation in stationary phase may be perceived a mild stress \cite{manhart2018trade}. These observations support our hypothesis that quiescence, a response to the relatively mild stress of starvation, may provide a helpful step towards resistance to harsher stress.

Previous work has demonstrated that yeast populations exposed to repeated cycles of growth-freeze-thaw increase their survival from $2\%$ to $70\%$ over the course of $100$ generations \cite{tvishamayi2024convergent}. The evolved strains show key characteristics of quiescent cells including smaller, denser cells, increased cell wall thickness. Our model captures this evolutionary trajectory qualitatively at the phenotypic level. It provides a framework to explore its dependence on the frequency and strength of environmental change, as well as on features of the population such as its maximum growth rate and the nature of the variation of lag time with the population phenotype.

Our model assumes that the growth rate, probability of quiescence, probability of survival as well as lag time of each cell depends on a single phenotype. In the case of yeast, we propose that this phenotype corresponds to the amount of intracellular trehalose produced by the cell once it reaches stationary phase. Previous work with \textit{E. coli} shows that phenotypic heterogeneity in lag times could be a strategy to break the limitations of the growth-survival trade-off \cite{moreno2020wide}. However, here we take a more holistic view by factoring in both the switch to quiescence, the exit from quiescence (through lag time) and their interaction with growth through dependence on the same phenotype. We take the functional form of this dependence to be the simplest assumptions that are biologically realistic; linear in the case of lag time, quadratic for growth rate and sigmoidal for the two probabilities. We are not aware of any attempt to quantify the dependence of these characteristics upon trehalose concentrations in yeast populations experimentally. Our results are nevertheless robust to changes in these functional forms. They depend only on the presence of a trade-off in the form of lag time and growth rate decreasing while the probability of survival and quiescence increase with the phenotype. Our results also are characteristic of selection from a threshold like selection pressure, and we expect similar behavior in many systems that show such a feature \cite{erdougan2024neutral, bershtein2017bridging, moorad2008levels, krebs1997estimating}.

We draw mutational fitness effects from a long-tailed Pearson distribution. The model is agnostic to the source of such mutations, which could be genetic or epigenetic in nature. Further, our framework does not speculate on the mechanism through which the memory of stress, or of intracellular trehalose produced in stationary phase can be inherited across generations. Interestingly, the whole genome sequencing of the evolved yeast strains in \cite{tvishamayi2024convergent} does not show any mutations conserved across replicate lines. This is also supported by previous work \cite{berry2011multiple}. Yet, a convergent phenotype emerges within the short period of $25$ generations, suggesting an epigenetic stress response. This response is still inheritable, as evidenced by the evolved phenotype persisting even in the absence of continued stress. It is possible that if quiescence, a generalized stress response mechanism, facilitates the adaptation to freeze thaw, that the adaptation mechanism is one that involves genetic assimilation \cite{badyaev2005stress, nanjundiah2003phenotypic}.  

We assume a stochastic switch to quiescence, with the probability of quiescence increasing with increasing intracellular trehalose. In this way our system is similar to stochastic switching to a persistent state in bacterial populations, although for us the probability of switching is biased by trehalose concentration in stationary phase \cite{wood2013bacterial, kussell2005bacterial, balaban2004bacterial}. However, our model is distinct from most models of bacterial persistors since we postulate that our phenotype is inheritable and over time lends higher survival fractions. There is some support for the idea that that persistence frequency in bacteria is also an evolvable trait that adapts to both the duration and frequency of the antibiotic
application\cite{wakamoto2013dynamic, fridman2014optimization}. We assume that the subset of quiescent cells is then subject to probabilistic survival upon exposure to stress. This is a reasonable assumption given previous evidence that quiescence leads to higher stress tolerance \cite{opalek2023systematic}. Yet this assumption remains experimentally unconfirmed, and it is not certain whether the same quiescent subset of the population in stationary phase is selected from by subsequent stress. 

Although the model presented here allows for a more sophisticated treatment of the biological processes involved in adaptation, we found that a simpler toy framework is also sufficient to capture the dependence of equilibrium survival on trade-off strength. Details of such a simpler framework which relies on a growth and survival fitness function is presented in Supplementary Information. In our model the trade-off strength is determined by the interplay between growth rate, quiescence and survival probabilities and lag time. In the toy model it is a more direct feature, the distance between the two fitness optima and selection strengths of the fitness functions.

We calculate the long term growth rate of a population adapting to growth-stress cycles and use it to calculate the optimum phenotype for different evolution programs. Our analysis reveals that the duration and nature of the growth process between successive stress exposures directly affects the optimum phenotype, final survival, as well as the long term growth rate. We compare the fitness of populations that are optimum under solely growth cycles and under growth-stress cycles. We calculate the long term growth rate of these 'growth-type' and 'stress-type' populations both with and without stress. Depending on the evolution program, the populations can evolve to be specialists or generalists. In particular, the long term growth rates of some stress-type populations remain positive and comparable to the long term growth rates of growth-type populations, both with and without stress. Since the long term growth rate of both strains is positive and comparable, this suggests the possibility of coexistence. We test this possibility using simulations of competition between the two strains under a similar growth paradigm with different growth durations, and indeed find coexistence under certain conditions.  This indicates that the overall fitness of the two strains does not show a strong trade-off despite the underlying growth-survival trade-off. Our simulations treat populations growing in a fixed finite resource, and this makes a direct comparison with the long term growth rate calculation, which relies on fixed finite time growth, difficult.

For a smaller range of lower values of $g_0$ and time of growth, $T$, we find that stress-type populations have positive long term growth rates in both types of environments, while growth-type populations do not survive cycles with stress exposure. Thus, for some evolution paradigms, populations do not show a significant decrease in their fitness in growth-only cycles while still increasing their fitness in cycles with stress exposure. This is because the overall fitness and long term growth rate depend not only on the relative strengths of the growth and survival functions but also on other characteristics of growth such as time, $T$ and maximum growth rate $g_0$. Several experimental studies have shown that fitness trade-offs are not ubiquitous in microbial systems \cite{bennett2007experimental, velicer1999evolutionary}. Our results demonstrate that the absence of more obvious fitness trade-offs does not preclude the existence of an underlying growth-survival trade-off. 

Although previous work has suggested that coexistence and trade-offs at the fitness level are context dependent \cite{pekkonen2013resource}, our calculations allow for a more systematic treatment of the different cases possible. Our results from calculations of the long term growth rate can easily be verified by evolving populations under different evolution programs, with varying growth times for instance. 

Our model is also reminiscent of recent work done on the population dynamics of bacteria in stochastic feast-famine cycles \cite{himeoka2020dynamics,niimi2025population}. Previous work which models growth and death processes both with growth-yield and growth-death trade offs shows that the ratio of the growth rate and death rate is the effective fitness of the population \cite{himeoka2020dynamics}. This is similar to our expression for the long term growth rate which combines survival probability with the growth rate. Further, they also find the duration of the feast and famine periods to be important to the eventual fate of the population. However, here we also take into account the added complexity of lag time along with growth and survival, all determined by a single phenotype. Other work has examined the optimal persistence strategy for bacteria in feast-famine cycles with the stochastic application of antibiotics \cite{laastad2025triggered} to find that 'triggered' dormancy, such as what we have included in this model with quiescence, may be the optimal strategy when the stress of the antibiotic treatment is very severe.





\begin{thebibliography}{10}

\bibitem{wenger2011hunger}
Jared~W Wenger, Jeffrey Piotrowski, Saisubramanian Nagarajan, Kami Chiotti, Gavin Sherlock, and Frank Rosenzweig.
\newblock Hunger artists: yeast adapted to carbon limitation show trade-offs under carbon sufficiency.
\newblock {\em PLoS genetics}, 7(8):e1002202, 2011.

\bibitem{zakrzewska2011genome}
Anna Zakrzewska, Gerco Van~Eikenhorst, Johanna~EC Burggraaff, Daniel~J Vis, Huub Hoefsloot, Daniela Delneri, Stephen~G Oliver, Stanley Brul, and Gertien~J Smits.
\newblock Genome-wide analysis of yeast stress survival and tolerance acquisition to analyze the central trade-off between growth rate and cellular robustness.
\newblock {\em Molecular biology of the cell}, 22(22):4435--4446, 2011.

\bibitem{shabestary2024phenotypic}
Kiyan Shabestary, Cinzia Klemm, Benedict Carling, James Marshall, Juline Savigny, Marko Storch, and Rodrigo Ledesma-Amaro.
\newblock Phenotypic heterogeneity follows a growth-viability tradeoff in response to amino acid identity.
\newblock {\em Nature Communications}, 15(1):6515, 2024.

\bibitem{bennett2007experimental}
Albert~F Bennett and Richard~E Lenski.
\newblock An experimental test of evolutionary trade-offs during temperature adaptation.
\newblock {\em Proceedings of the National Academy of Sciences}, 104(suppl\_1):8649--8654, 2007.

\bibitem{buckling2007experimental}
Angus Buckling, Michael~A Brockhurst, Michael Travisano, and Paul~B Rainey.
\newblock Experimental adaptation to high and low quality environments under different scales of temporal variation.
\newblock {\em Journal of Evolutionary Biology}, 20(1):296--300, 2007.

\bibitem{lambros2021emerging}
Maryl Lambros, Ximo Pechuan-Jorge, Daniel Biro, Kenny Ye, and Aviv Bergman.
\newblock Emerging adaptive strategies under temperature fluctuations in a laboratory evolution experiment of escherichia coli.
\newblock {\em Frontiers in microbiology}, 12:724982, 2021.

\bibitem{lopez2008tuning}
Luis L{\'o}pez-Maury, Samuel Marguerat, and J{\"u}rg B{\"a}hler.
\newblock Tuning gene expression to changing environments: from rapid responses to evolutionary adaptation.
\newblock {\em Nature Reviews Genetics}, 9(8):583--593, 2008.

\bibitem{tikhonov2020model}
Mikhail Tikhonov, Shamit Kachru, and Daniel~S Fisher.
\newblock A model for the interplay between plastic tradeoffs and evolution in changing environments.
\newblock {\em Proceedings of the National Academy of Sciences}, 117(16):8934--8940, 2020.

\bibitem{tvishamayi2024convergent}
Charuhansini Tvishamayi, Farhan Ali, Nandita Chaturvedi, Nithila Madhu-Kumar, Zeenat Rashida, Chandan~Muni Reddy, Ankita Ray, Stephan Herminghaus, and Shashi Thutupalli.
\newblock Convergent cellular adaptation to freeze-thaw stress via a quiescence-like state in yeast.
\newblock {\em eLife}, July 2025.

\bibitem{olsson2020structural}
Christoffer Olsson and Jan Swenson.
\newblock Structural comparison between sucrose and trehalose in aqueous solution.
\newblock {\em The Journal of Physical Chemistry B}, 124(15):3074--3082, 2020.

\bibitem{aranda2004trehalose}
Juan~S Aranda, Edgar Salgado, and Patricia Taillandier.
\newblock Trehalose accumulation in saccharomyces cerevisiae cells: experimental data and structured modeling.
\newblock {\em Biochemical Engineering Journal}, 17(2):129--140, 2004.

\bibitem{tapia2015increasing}
Hugo Tapia, Lindsey Young, Douglas Fox, Carolyn~R Bertozzi, and Douglas Koshland.
\newblock Increasing intracellular trehalose is sufficient to confer desiccation tolerance to saccharomyces cerevisiae.
\newblock {\em Proceedings of the National Academy of Sciences}, 112(19):6122--6127, 2015.

\bibitem{arguelles2000physiological}
Juan~Carlos Arg{\"u}elles.
\newblock Physiological roles of trehalose in bacteria and yeasts: a comparative analysis.
\newblock {\em Archives of microbiology}, 174:217--224, 2000.

\bibitem{panek1963function}
Anita Panek.
\newblock Function of trehalose in baker's yeast (saccharomyces cerevisiae).
\newblock {\em Archives of Biochemistry and Biophysics}, 100(3):422--425, 1963.

\bibitem{diniz1999preservation}
L~Diniz-Mendes, E~Bernardes, Pedro~Soares De~Araujo, AD~Panek, and VMF Paschoalin.
\newblock Preservation of frozen yeast cells by trehalose.
\newblock {\em Biotechnology and Bioengineering}, 65(5):572--578, 1999.

\bibitem{shi2010trehalose}
Lei Shi, Benjamin~M Sutter, Xinyue Ye, and Benjamin~P Tu.
\newblock Trehalose is a key determinant of the quiescent metabolic state that fuels cell cycle progression upon return to growth.
\newblock {\em Molecular biology of the cell}, 21(12):1982--1990, 2010.

\bibitem{opalek2022ecological}
Monika Opalek, Bogna Smug, Michael Doebeli, and Dominika Wloch-Salamon.
\newblock On the ecological significance of phenotypic heterogeneity in microbial populations undergoing starvation.
\newblock {\em Microbiology Spectrum}, 10(1):e00450--21, 2022.

\bibitem{chu2016lag}
Dominique Chu and David~J Barnes.
\newblock The lag-phase during diauxic growth is a trade-off between fast adaptation and high growth rate.
\newblock {\em Scientific reports}, 6(1):25191, 2016.

\bibitem{basan2020universal}
Markus Basan, Tomoya Honda, Dimitris Christodoulou, Manuel H{\"o}rl, Yu-Fang Chang, Emanuele Leoncini, Avik Mukherjee, Hiroyuki Okano, Brian~R Taylor, Josh~M Silverman, et~al.
\newblock A universal trade-off between growth and lag in fluctuating environments.
\newblock {\em Nature}, 584(7821):470--474, 2020.

\bibitem{stern2007genome}
Shay Stern, Tali Dror, Elad Stolovicki, Naama Brenner, and Erez Braun.
\newblock Genome-wide transcriptional plasticity underlies cellular adaptation to novel challenge.
\newblock {\em Molecular Systems Biology}, 3(1):106, 2007.

\bibitem{soontorngun2017reprogramming}
Nitnipa Soontorngun.
\newblock Reprogramming of nonfermentative metabolism by stress-responsive transcription factors in the yeast saccharomyces cerevisiae.
\newblock {\em Current genetics}, 63:1--7, 2017.

\bibitem{de2012essence}
Claudio De~Virgilio.
\newblock The essence of yeast quiescence.
\newblock {\em FEMS microbiology reviews}, 36(2):306--339, 2012.

\bibitem{sagot2019cell}
Isabelle Sagot and Damien Laporte.
\newblock The cell biology of quiescent yeast--a diversity of individual scenarios.
\newblock {\em Journal of Cell Science}, 132(1):jcs213025, 2019.

\bibitem{breeden2022quiescence}
Linda~L Breeden and Toshio Tsukiyama.
\newblock Quiescence in saccharomyces cerevisiae.
\newblock {\em Annual Review of Genetics}, 56(1):253--278, 2022.

\bibitem{aragon2008characterization}
Anthony~D Aragon, Angelina~L Rodriguez, Osorio Meirelles, Sushmita Roy, George~S Davidson, Phillip~H Tapia, Chris Allen, Ray Joe, Don Benn, and Margaret Werner-Washburne.
\newblock Characterization of differentiated quiescent and nonquiescent cells in yeast stationary-phase cultures.
\newblock {\em Molecular biology of the cell}, 19(3):1271--1280, 2008.

\bibitem{daignan2011proliferation}
Bertrand Daignan-Fornier and Isabelle Sagot.
\newblock Proliferation/quiescence: the controversial" aller-retour".
\newblock {\em Cell Division}, 6:1--4, 2011.

\bibitem{valcourt2012staying}
James~R Valcourt, Johanna~MS Lemons, Erin~M Haley, Mina Kojima, Olukunle~O Demuren, and Hilary~A Coller.
\newblock Staying alive: metabolic adaptations to quiescence.
\newblock {\em Cell cycle}, 11(9):1680--1696, 2012.

\bibitem{herman2002stationary}
Paul~K Herman.
\newblock Stationary phase in yeast.
\newblock {\em Current opinion in microbiology}, 5(6):602--607, 2002.

\bibitem{allen2006isolation}
Chris Allen, Sabrina B{\"u}ttner, Anthony~D Aragon, Jason~A Thomas, Osorio Meirelles, Jason~E Jaetao, Don Benn, Stephanie~W Ruby, Marten Veenhuis, Frank Madeo, et~al.
\newblock Isolation of quiescent and nonquiescent cells from yeast stationary-phase cultures.
\newblock {\em The Journal of cell biology}, 174(1):89--100, 2006.

\bibitem{werner1993stationary}
Margaret Werner-Washburne, Edward Braun, Gerald~C Johnston, and Richard~A Singer.
\newblock Stationary phase in the yeast saccharomyces cerevisiae.
\newblock {\em Microbiological reviews}, 57(2):383--401, 1993.

\bibitem{alarcon2011quiescence}
Tom{\'a}s Alarc{\'o}n and Henrik~Jeldtoft Jensen.
\newblock Quiescence: a mechanism for escaping the effects of drug on cell populations.
\newblock {\em Journal of The Royal Society Interface}, 8(54):99--106, 2011.

\bibitem{wang2017exit}
Xia Wang, Kotaro Fujimaki, Geoffrey~C Mitchell, Jungeun~Sarah Kwon, Kimiko Della~Croce, Chris Langsdorf, Hao~Helen Zhang, and Guang Yao.
\newblock Exit from quiescence displays a memory of cell growth and division.
\newblock {\em Nature communications}, 8(1):321, 2017.

\bibitem{sun2021cellular}
Siyu Sun and David Gresham.
\newblock Cellular quiescence in budding yeast.
\newblock {\em Yeast}, 38(1):12--29, 2021.

\bibitem{kaplan2021observation}
Yoav Kaplan, Shaked Reich, Elyaqim Oster, Shani Maoz, Irit Levin-Reisman, Irine Ronin, Orit Gefen, Oded Agam, and Nathalie~Q Balaban.
\newblock Observation of universal ageing dynamics in antibiotic persistence.
\newblock {\em Nature}, 600(7888):290--294, 2021.

\bibitem{woronoff2020metabolic}
Gabrielle Woronoff, Philippe Nghe, Jean Baudry, Laurent Boitard, Erez Braun, Andrew~D Griffiths, and J{\'e}r{\^o}me Bibette.
\newblock Metabolic cost of rapid adaptation of single yeast cells.
\newblock {\em Proceedings of the National Academy of Sciences}, 117(20):10660--10666, 2020.

\bibitem{zhu2024shaping}
Manlu Zhu and Xiongfeng Dai.
\newblock Shaping of microbial phenotypes by trade-offs.
\newblock {\em Nature Communications}, 15(1):4238, 2024.

\bibitem{bruggeman2023trade}
Frank~J Bruggeman, Bas Teusink, and Ralf Steuer.
\newblock Trade-offs between the instantaneous growth rate and long-term fitness: consequences for microbial physiology and predictive computational models.
\newblock {\em Bioessays}, 45(10):2300015, 2023.

\bibitem{jahn2017adaptive}
Leonie~J Jahn, Christian Munck, Mostafa~MH Ellabaan, and Morten~OA Sommer.
\newblock Adaptive laboratory evolution of antibiotic resistance using different selection regimes lead to similar phenotypes and genotypes.
\newblock {\em Frontiers in microbiology}, 8:816, 2017.

\bibitem{hughes2012selection}
Diarmaid Hughes and Dan~I Andersson.
\newblock Selection of resistance at lethal and non-lethal antibiotic concentrations.
\newblock {\em Current opinion in microbiology}, 15(5):555--560, 2012.

\bibitem{sandegren2014selection}
Linus Sandegren.
\newblock Selection of antibiotic resistance at very low antibiotic concentrations.
\newblock {\em Upsala journal of medical sciences}, 119(2):103--107, 2014.

\bibitem{manhart2018trade}
Michael Manhart, Bharat~V Adkar, and Eugene~I Shakhnovich.
\newblock Trade-offs between microbial growth phases lead to frequency-dependent and non-transitive selection.
\newblock {\em Proceedings of the Royal Society B: Biological Sciences}, 285(1872):20172459, 2018.

\bibitem{moreno2020wide}
Stefany Moreno-G{\'a}mez, Daniel~J Kiviet, Cl{\'e}ment Vulin, Susan Schlegel, Kim Schlegel, G~Sander van Doorn, and Martin Ackermann.
\newblock Wide lag time distributions break a trade-off between reproduction and survival in bacteria.
\newblock {\em Proceedings of the National Academy of Sciences}, 117(31):18729--18736, 2020.

\bibitem{erdougan2024neutral}
Ay{\c{s}}e~Nisan Erdo{\u{g}}an, Pouria Dasmeh, Raymond~D Socha, John~Z Chen, Benjamin~E Life, Rachel Jun, Linda Kiritchkov, Dan Kehila, Adrian~WR Serohijos, and Nobuhiko Tokuriki.
\newblock Neutral drift upon threshold-like selection promotes variation in antibiotic resistance phenotype.
\newblock {\em Nature Communications}, 15(1):10813, 2024.

\bibitem{bershtein2017bridging}
Shimon Bershtein, Adrian~WR Serohijos, and Eugene~I Shakhnovich.
\newblock Bridging the physical scales in evolutionary biology: from protein sequence space to fitness of organisms and populations.
\newblock {\em Current opinion in structural biology}, 42:31--40, 2017.

\bibitem{moorad2008levels}
Jacob~A Moorad and Timothy~A Linksvayer.
\newblock Levels of selection on threshold characters.
\newblock {\em Genetics}, 179(2):899--905, 2008.

\bibitem{krebs1997estimating}
Robert~A Krebs and Volker Loeschcke.
\newblock Estimating heritability in a threshold trait: heat-shock tolerance in drosophila buzzatii.
\newblock {\em Heredity}, 79(3):252--259, 1997.

\bibitem{berry2011multiple}
David~B Berry, Qiaoning Guan, James Hose, Suraiya Haroon, Marinella Gebbia, Lawrence~E Heisler, Corey Nislow, Guri Giaever, and Audrey~P Gasch.
\newblock Multiple means to the same end: the genetic basis of acquired stress resistance in yeast.
\newblock {\em PLoS genetics}, 7(11):e1002353, 2011.

\bibitem{badyaev2005stress}
Alexander~V Badyaev.
\newblock Stress-induced variation in evolution: from behavioural plasticity to genetic assimilation.
\newblock {\em Proceedings of the Royal Society B: Biological Sciences}, 272(1566):877--886, 2005.

\bibitem{nanjundiah2003phenotypic}
Vidyanand Nanjundiah.
\newblock Phenotypic plasticity and evolution by genetic assimilation.
\newblock {\em Origination of organismal form: Beyond the gene in developmental and evolutionary biology}, 2:245, 2003.

\bibitem{wood2013bacterial}
Thomas~K Wood, Stephen~J Knabel, and Brian~W Kwan.
\newblock Bacterial persister cell formation and dormancy.
\newblock {\em Applied and environmental microbiology}, 79(23):7116--7121, 2013.

\bibitem{kussell2005bacterial}
Edo Kussell, Roy Kishony, Nathalie~Q Balaban, and Stanislas Leibler.
\newblock Bacterial persistence: a model of survival in changing environments.
\newblock {\em Genetics}, 169(4):1807--1814, 2005.

\bibitem{balaban2004bacterial}
Nathalie~Q Balaban, Jack Merrin, Remy Chait, Lukasz Kowalik, and Stanislas Leibler.
\newblock Bacterial persistence as a phenotypic switch.
\newblock {\em Science}, 305(5690):1622--1625, 2004.

\bibitem{van2016frequency}
Bram Van~den Bergh, Joran~E Michiels, Tom Wenseleers, Etthel~M Windels, Pieterjan~Vanden Boer, Donaat Kestemont, Luc De~Meester, Kevin~J Verstrepen, Natalie Verstraeten, Maarten Fauvart, et~al.
\newblock Frequency of antibiotic application drives rapid evolutionary adaptation of escherichia coli persistence.
\newblock {\em Nature microbiology}, 1(5):1--7, 2016.

\bibitem{fridman2014optimization}
Ofer Fridman, Amir Goldberg, Irine Ronin, Noam Shoresh, and Nathalie~Q Balaban.
\newblock Optimization of lag time underlies antibiotic tolerance in evolved bacterial populations.
\newblock {\em Nature}, 513(7518):418--421, 2014.

\bibitem{opalek2023systematic}
Monika Opalek, Hanna Tutaj, Adrian Pirog, Bogna~J Smug, Joanna Rutkowska, and Dominika Wloch-Salamon.
\newblock A systematic review on quiescent state research approaches in s. cerevisiae.
\newblock {\em Cells}, 12(12):1608, 2023.

\bibitem{velicer1999evolutionary}
Gregory~J Velicer and Richard~E Lenski.
\newblock Evolutionary trade-offs under conditions of resource abundance and scarcity: experiments with bacteria.
\newblock {\em Ecology}, 80(4):1168--1179, 1999.

\bibitem{pekkonen2013resource}
Minna Pekkonen, Tarmo Ketola, and Jouni~T Laakso.
\newblock Resource availability and competition shape the evolution of survival and growth ability in a bacterial community.
\newblock {\em PLoS One}, 8(9):e76471, 2013.

\bibitem{himeoka2020dynamics}
Yusuke Himeoka and Namiko Mitarai.
\newblock Dynamics of bacterial populations under the feast-famine cycles.
\newblock {\em Physical Review Research}, 2(1):013372, 2020.

\bibitem{niimi2025population}
Rintaro Niimi, Chikara Furusawa, and Yusuke Himeoka.
\newblock Population dynamics of generalist/specialist strategies in the feast-famine cycle.
\newblock {\em bioRxiv}, pages 2025--05, 2025.

\bibitem{laastad2025triggered}
Silja~Borring L{\aa}stad and Namiko Mitarai.
\newblock Triggered and spontaneous dormancy in bacteria during feast-famine cycles with stochastic antibiotic application.
\newblock {\em bioRxiv}, pages 2025--02, 2025.

\bibitem{wakamoto2013dynamic}
Yuichi~Wakamoto, Neeraj~Dhar, Remy~Chait, Katrin~Schneider, Fran{\c{c}}ois~Signorino-Gelo, Stanislas~Leibler, and John~D.~McKinney.
\newblock Dynamic persistence of antibiotic-stressed mycobacteria.
\newblock {\em Science}, 339(6115):91--95, 2013.

\bibitem{sagot2019quiescence}
Isabelle Sagot and Damien Laporte.
\newblock Quiescence, an individual journey.
\newblock {\em Current genetics}, 65(3):695--699, 2019.


\end{thebibliography}
\end{document}